\documentclass[electronics,article,submit,pdftex,moreauthors]{Definitions/mdpi}

\renewcommand{\linenumbers}{}

\firstpage{1} 
\pubvolume{–}
\issuenum{–}
\articlenumber{–}
\pubyear{2024}
\copyrightyear{2024}
\datereceived{–} 
\dateaccepted{–} 
\datepublished{–} 
\hreflink{https://doi.org/} % If needed use \linebreak
\usepackage[frozencache,cachedir=.]{minted}
\usepackage{subcaption}

\makeatletter
\AtBeginEnvironment{minted}{\dontdofcolorbox}
\def\dontdofcolorbox{\renewcommand\fcolorbox[4][]{##4}}
\makeatother

\Title{RDF Stream Taxonomy: Systematizing RDF Stream Types in Research and Practice}

\TitleCitation{RDF Stream Taxonomy: Systematizing RDF Stream Types in Research and Practice}

\Author{Piotr Sowiński $^{1,2,}$*\orcidA{}, Paweł Szmeja $^{2}$\orcidB{}, Maria Ganzha $^{1,2}$\orcidB{} and Marcin Paprzycki $^{2}$\orcidD{}}

\AuthorNames{Piotr Sowiński, Paweł Szmeja, Maria Ganzha and Marcin Paprzycki}

\AuthorCitation{Sowiński, P.; Szmeja, P.; Ganzha, M.; Paprzycki, M.}

\address{%
$^{1}$ \quad 
Faculty of Mathematics and Information Science, Warsaw University of Technology, ul. Koszykowa 75, 00-662 Warsaw, Poland\\
$^{2}$ \quad 
Systems Research Institute, Polish Academy of Sciences, ul. Newelska 6, 01-447 Warsaw, Poland}

\corres{Correspondence: piotr.sowinski@ibspan.waw.pl}

\abstract{Over the years, RDF streaming was explored in research and practice from many angles, resulting in a wide range of RDF stream definitions. This variety presents a major challenge in discussing and integrating streaming systems, due to the lack of a common language. This work attempts to address this critical research gap, by systematizing RDF stream types present in the literature in a novel taxonomy. The proposed RDF Stream Taxonomy (RDF-STaX) is embodied in an OWL 2 DL ontology that follows the FAIR principles, making it readily applicable in practice. Extensive documentation and additional resources are provided, to foster the adoption of the ontology. Three use cases for the ontology are presented with accompanying competency questions, demonstrating the usefulness of the resource. Additionally, this work introduces a novel nanopublications dataset, which serves as a collaborative, living state-of-the-art review of RDF streaming. The results of a multifaceted evaluation of the resource are presented, testing its logical validity, use case coverage, and adherence to the community’s best practices, while also comparing it to other works. RDF-STaX is expected to help drive innovation in RDF streaming, by fostering scientific discussion, cooperation, and tool interoperability.}


\begin{document}

%%%%%%%%%%%%%%%%%%%%%%%%%%%%%%%%%%%%%%%%%%
\section{Introduction}

The Resource Description Framework (RDF) is a standard-backed foundation for modeling and exchanging knowledge graphs, with a wide range of practical applications~\cite{pan2009resource, hitzler2021review}. The RDF model is based on the building blocks of the Web (e.g., it uses Internationalized Resource Identifiers for node names), making it very flexible and extensible~\cite{Cyganiak:14:RCA}. The list of use cases for RDF has expanded greatly since its inception, with time moving beyond applications where the modeled knowledge is treated as a single, static dataset. More recently, the concept of RDF streams has gained traction in research and industrial applications. Generally, an RDF stream refers to a potentially infinite stream of data modeled using the RDF standard, to which stream processing principles apply~\cite{kleppmann2017designing}. This research interest resulted in numerous contributions to the field, such as streaming serializations, querying and reasoning engines, programming frameworks, datasets, and other resources~\cite{bonte2023streaming, sowinski2022efficient, tommasini2023streaming, van2023declarative}. Without a doubt, streaming RDF solutions are in demand, and more innovation is still needed. However, the landscape of RDF streaming terminology is fragmented, with many dissimilar streaming task formulations and incompatible solutions.

The term ``RDF stream'' is used in the literature to refer to many different concepts, ranging from theoretical models~\cite{dell2014rsp}, to very technical ones, focusing on the representation and practical usage~\cite{fernandez2014efficient, oo2022rmlstreamer, sowinski2022efficient}. As noted in past research works~\cite{dell2016query, le2009rapid, schraudner2022stream}, RDF streaming systems need more standardization in stream exchange protocols and stream task definitions. This would be beneficial not only in the academic context, but would also facilitate the distribution and reuse of RDF streams in practice. Additionally, it could enable streaming tool interoperability through the inclusion of appropriate metadata. At the same time, if the research is to solve real-world problems, the variety of existing RDF streaming task formulations must be embraced, not limiting oneself to one specific type of RDF streams. However, up until now, there was very little academic discussion about the existing RDF stream types -- what stream types are used, how they relate to each other, and which use cases are they applicable to. Our work aims to address this critical research gap, which has a large impact on the uptake of RDF streaming in general.

The main contribution of this work is the RDF Stream Taxonomy (RDF-STaX), embodied in an OWL 2 DL ontology. This novel taxonomy systematizes the existing RDF stream types on the basis of a broad state-of-the-art review. RDF-STaX aims to be a practical and useful resource for researchers and practitioners, enabling them to semantically annotate RDF stream types in publications, software, and streams published on the Web. To achieve this, RDF-STaX follows the community's best practices for ontology publishing and is accompanied with extensive documentation, including usage examples. The three envisioned use cases for RDF-STaX are presented with accompanying competency questions implemented in SPARQL. Two of these use cases were already implemented in practice -- a living state-of-the-art review of RDF streaming using nanopublications, and annotating stream types in RiverBench datasets. Finally, RDF-STaX is compared to other ontologies, highlighting the gap which it covers.

This work is organized as follows. Section~\ref{sec:literature} presents a comprehensive literature review of the topic. Section~\ref{sec:systematize} introduces the proposed systematization of RDF streams, along with an alignment of these definitions to the state of the art. Section~\ref{sec:onto} describes the proposed RDF-STaX ontology. Section~\ref{sec:app} presents the use cases envisioned for RDF-STaX, while Section~\ref{sec:evaluation} includes a multifaceted evaluation of the proposed ontology, including a comparison to other ontologies. Finally, Section~\ref{sec:discussion} discusses the results and outlines future work directions, with Section~\ref{sec:conclusion} presenting the concluding remarks.

\section{Literature Review}
\label{sec:literature}

The primary goal of this paper is to introduce a taxonomy of RDF stream types, which naturally must be grounded in the state of the art to be relevant. As no comprehensive and up-to-date review of this topic is available, here we attempt to summarize the most significant works that use the concept of RDF streams.

The scope of this study only includes RDF streams that can be fully represented using the RDF 1.1 W3C Recommendation~\cite{Cyganiak:14:RCA}. This notably excludes theoretical stream models that do not have a direct translation to RDF, such as RSP-QL~\cite{dell2014rsp}. However, due to the importance of such models, they are briefly discussed to provide additional context. The reason to not consider purely theoretical models is that they are not directly comparable to models that have a fixed representation in RDF 1.1, as they operate on a different level of abstraction. Therefore, including both purely theoretical and fully representable RDF streams in the review would make the resulting taxonomy methodologically flawed. This scope decision is additionally motivated by the fact that only streams fully representable in RDF can be easily exchanged and published on the Web using already existing protocols, while the theoretical models have more limited applicability, for example, only within one specific system (e.g., a stream reasoning engine). This choice is thus natural due this work's focus on solving the challenge of streaming tool interoperability and harmonizing the ways in which RDF streams are exchanged.

To cover the various areas of research related to RDF streaming, several survey papers and books were included in the review~\cite{bonte2023streaming, keskisarkka2013semantic, llanes2016sensor, ma2016storing, modoni2014survey, ozsu2016survey, su2016stream, tommasini2023streaming, van2023declarative, zhang2021rdf}, on topics such as RDF data management, linked streaming data, and stream reasoning. 
For many of the examined works, determining the exact nature of the stream was not trivial, due to unclear descriptions and a lack of standardized terminology. Therefore, not only the papers were examined, but also the code and documentation, where available.

\subsection{RDF Streaming Protocols}

One prominent area of research that uses RDF streams are works focusing on streaming protocols for transmitting RDF data. One of the earlier examples is Streaming HDT (S-HDT)~\cite{hasemann2012shdt}, which facilitates streaming RDF data in IoT devices. More specifically, in S-HDT producers (IoT devices) emit streams of compressed, unnamed RDF graphs. RDSZ~\cite{fernandez2014rdsz} is a later method that also attempts streaming compressed RDF graphs over the network. ERI~\cite{fernandez2014efficient} is a similar approach, however, it has several views of the RDF stream. At the highest level, ERI streams a continuous sequence of triples that is then divided into discrete blocks (RDF graphs). Then, on the lowest level, these graphs are decomposed into \emph{subject-molecules} (graphs with a single RDF subject) that are then streamed over the network. Therefore, ERI has three different views of the streaming problem, with each corresponding to a different level in the protocol.

The RDF EXI communication protocol for constrained devices~\cite{kabisch2015exi} focuses on streaming RDF graphs in a very efficient manner. Finally, the recent Jelly protocol~\cite{sowinski2022efficient} treats RDF streams similarly to ERI, with it streaming a sequence of RDF statements divided into discrete elements (stream frames). The main difference is that Jelly can also stream quad statements, making it generalizable to RDF datasets. It should be noted here that none of these works assume time annotations to be a necessary part of RDF streams, in contrast to the works described in the following subsection.

\subsection{Stream Processing and Reasoning}

There is a large body of research on the topics of RDF stream reasoning and stream processing, where a special focus is often placed on the time annotations of stream elements. We could find only one work in this category that did not consider the temporal aspect, an early proposal for a streaming SPARQL engine from 2007 by Groppe et al.~\cite{groppe2007sparql}. The proposed engine executes queries over streams of RDF triples, in real time.

Most publications in this category focus on the inner workings of querying/reasoning engines, and adopt task formulations that are not fully grounded in the RDF specification. The most popular such formulation is the stream of timestamped triples \cite{anicic2011ep, barbieri2010querying, bolles2008streaming, calbimonte2016evaluating, dell2014rsp, dell2015towards, komazec2012sparkwave, le2011native, tommasini2021rsp4j, zhang2021rdf}, where a stream is a sequence of $\langle triple, timestamp \rangle$ tuples. There are many variations of this task, with time intervals instead of timestamps~\cite{anicic2011ep}, multiple timestamps per triple~\cite{dell2013correctness, tommasini2021rsp4j}, and different time domains. However, this model (and other equivalent or similar models, e.g., time-varying graphs~\cite{dell2016query, le2015rdf, tommasini2021rsp4j}) by itself is not enough to fully realize and exchange an RDF stream -- this would require encoding the time information in RDF, which is often omitted in these works.

The lack of grounding in the RDF specification of the timestamped triple model is a known problem~\cite{calbimonte2016evaluating}, and solutions to it were proposed. The most prominent formulation is the stream of timestamped named graphs, where each stream element is a named RDF graph. The timestamp (or other metadata) of this graph is added by making statements in the default graph about the named graph's name node~\cite{calbimonte2017linked, mauri2016triplewave, wu2022workflow}. This representation was possibly first proposed by Tappolet \& Bernstein in 2009~\cite{tappolet2009applied}, and was then expanded by the W3C RSP Community Group, which created a draft specification for the RSP Data Model~\cite{rspmodel}. The RSP Data Model uses the aforementioned formulation to define RDF streams, but also mentions that RSP engines may consume streams of graphs, datasets, or named graphs and then turn them into RDF streams proper by associating each element with a timestamp. The RSP Data Model is the most mature and best-grounded model for representing timestamped streams, having been drafted by the community, and even proposing the formal semantics of RDF streams. However, it was never made into a W3C Recommendation, and the RSP Community Group at the time of this work's submission \href{https://lists.w3.org/Archives/Public/public-rsp/2023Mar/0000.html}{appears to be inactive} (accessed on 4 June 2024).

A similar representation for timestamped streams was proposed in 2010 by Barbieri \& Della Valle~\cite{barbieri2010proposal}. In the proposal, stream elements are named RDF graphs (called instantaneous graphs). These graphs are linked together with additional metadata placed in an additional named graph dedicated to this purpose (called a stream graph). Each instantaneous graphs' name node is associated with a timestamp triple using the \texttt{sld:receivedAt} property. This formulation differs from the RSP Data Model mostly by placing the metadata in a named graph instead of the default graph.

Time-Annotated RDF (TA-RDF)~\cite{rodriguez2009semantic} is an alternative RDF representation for the timestamped stream problem (more precisely, TA-RDF has a direct translation to RDF). 
It identifies each stream element as an URI node in an RDF graph. The data relevant to that stream element is then attached to the node. This boils down to each stream element being an RDF graph in which a single node serves as the ``point of entry''. This model is notably less flexible than the RSP Data Model, as it requires the entire content of each stream element to be connected to this single node. A similar approach is used by Stream Containers~\cite{schraudner2022stream}, a framework for publishing RDF streams on the Web. There, a stream is a sequence of unnamed RDF graphs that are referenced by URIs. This is not the same as RDF named graphs, as each URI instead points to a node (the element's subject) within one of the unnamed graphs, and the temporal information is also contained within this graph.

One auxiliary RDF stream definition also appears in this category. The VoCaLS vocabulary for describing linked streams~\cite{tommasini2018vocals} defines an RDF stream as a potentially infinite sequence of RDF graphs and/or triples. This definition is used in practice in the RSP4J API~\cite{tommasini2021rsp4j}, whose YASPER implementation can consume streams of RDF graphs or RDF triples.

\subsection{Streaming Semantic Annotation and Translation}

Several systems were developed for streaming semantic annotation (lifting) and translation. RMLStreamer~\cite{haesendonck2019parallel, oo2022rmlstreamer} annotates data using the RML language~\cite{dimou2014rml}, outputting a stream of RDF datasets. CARML~\cite{carml} is a Java library that provides generic facilities for annotating data using RML. It supports outputting a stream of RDF quad statements (using RDF4J Rio~\cite{rio}), which corresponds to a single RDF dataset. CARML does not natively support streams of datasets, and thus it operates on a different abstraction level than RMLStreamer. This limitation is addressed by the Semantic Annotation enabler~\cite{semann} from the ASSIST-IoT project~\cite{szmeja2023assist}, which outputs a stream of RDF datasets, each corresponding to one input document. SPARQL-Generate~\cite{lefranccois2016flexible} is an extension of the SPARQL language, which allows for generating RDF from various data sources. Its \href{https://github.com/sparql-generate/sparql-generate}{reference implementation} (accessed on 4 June 2024) supports outputting a stream of RDF triples (using Apache Jena RIOT~\cite{riot}). 

Last in this category is the Inter Platform Semantic Mediator (IPSM)~\cite{ganzha2017streaming}, which translates a stream of RDF graphs into another such stream, in real time. Although on a logical level IPSM processes RDF graphs, it physically outputs a stream of RDF datasets, where each dataset has two named graphs -- one containing the translated RDF, and the other with metadata about the translation itself.

\subsection{Semantic Streaming Applications}

Over the years, RDF streams were employed in various real-life use cases. One such example is DBpedia-Live~\cite{morsey2012dbpedia} -- a now-defunct service that published changes to the DBpedia knowledge base in real time. Two streams were published -- deleted and added triples. In each stream, the elements were RDF graphs, in the form of N-Triples files.

RDF streams are also commonly used in semantic sensor networks~\cite{llanes2016sensor}. In this context, the Graph of Things (GoT)~\cite{le2016graph} was proposed, which uses streams of RDF graphs. Another common use case is monitoring social media activity. Here, a survey by Keskis{\"a}rkk{\"a} \& Blomqvist~\cite{keskisarkka2013semantic} noted that RDF streams should have RDF graphs as elements, as this approach better addresses the needs of their use case. However, they also acknowledge that RDF streams can sometimes have RDF triples as elements. Streams are also used for sending updates between decentralized social network servers, as in the ActivityPub protocol~\cite{Tallon:18:A} standardized by the W3C. In ActivityPub, an item of activity is described with an RDF graph (in JSON-LD). These items form ordered streams of RDF graphs which are then exchanged between the nodes.
The Linked Data Event Streams (LDES)~\cite{van2021ldes} proposal addresses publishing streams composed of elements, where each element is a fragment of an RDF graph, identified by a specific subject node. Finally, the recent proposal for a live open scientific knowledge graph~\cite{le2022towards} uses streams of RDF graphs to monitor scientific and social activity in real time.

\subsection{Streaming I/O}

The last broad category of RDF streaming solutions are those that process simple sequences of RDF statements (triples or quads) for input/output (I/O) operations. Two examples are the RDF4J Rio toolkit~\cite{rio} and Apache Jena's RIOT~\cite{riot}, both of which can parse and serialize streams of statements, reading from or writing to a Java byte stream. The two most obvious serializations for this task are N-Triples and N-Quads, however, Jena also supports \emph{streamed block formats} for Turtle and Trig, where statements are transmitted in small batches (blocks).

\subsection{Summary}
\label{sec:review_summary}

The review uncovered a wide variety of task formulations for RDF streams. It should be noted that most of the identified works use the exact same term (RDF stream) to describe very different concepts. The review also highlighted an important distinction between theoretical models of RDF streams (e.g., timestamped triples) and practical ones, making it clear that direct comparisons between the two are not possible. This distinction was also observed earlier in several works~\cite{bonte2023streaming, calbimonte2016evaluating, dell2013correctness, schraudner2022stream}. Another contrasting feature is the use of graphs or datasets as stream elements, versus singular triple or quad statements. This was also noticed in earlier works focusing on the practical aspects of implementing RDF streams~\cite{bonte2023streaming, keskisarkka2013semantic}.

Worth noting is that some works have several views of the same RDF stream. For example, ERI~\cite{fernandez2014efficient} views its RDF streams as a flow of triples, blocks (RDF graphs), or subject-molecules (RDF subgraphs), depending on the level of abstraction. This suggests that describing the type of an RDF stream depends on the context in which it is considered, or, metaphorically, one's perspective. In fact, an N-Triples file would be considered by RDF4J's Rio~\cite{rio} as a stream of triples, while an application like DBpedia-Live~\cite{morsey2012dbpedia} or IPSM~\cite{ganzha2017streaming} would see it as a single element in a stream of graphs.

The diversity of problem statements encountered in the literature clearly suggests that expecting there to be a single model of RDF streams may be unrealistic, especially given that each stream may be understood in several different ways, depending on the perspective or context. Each of the identified approaches is motivated by some use case and thus is valid. Therefore, we must seek solutions to better understand and harmonize this task diversity, which is the primary goal of RDF-STaX.

\section{Proposed Systematization}
\label{sec:systematize}

In what follows, we propose a set of definitions for the types of RDF streams present in the literature and current practice. Then, Section~\ref{sec:taxo} organizes these definitions in the RDF Stream Taxonomy, and Section~\ref{sec:correspondence} links the proposed stream types to the state-of-the-art review performed in the previous section. The basis for all following definitions is the RDF 1.1 W3C Recommendation~\cite{Cyganiak:14:RCA}, which formally defines terms such as RDF graphs and RDF triples, and must be implemented by any RDF 1.1-compliant software. We use it with the aim of making our definitions readily applicable in the context of existing, RDF 1.1-compliant software.

The model-theoric semantics of RDF streams are not considered in this work. Instead, the proposed semi-formal definitions can be used as a universal basis on top of which the formal semantics can be defined, as needed. This is motivated by, firstly, the semantics of RDF streams being entirely irrelevant for some of the discussed application areas (e.g., streaming protocols, streaming I/O). Secondly, only the semantics of RDF graphs are well-defined -- there are no agreed-upon formal semantics for RDF datasets~\cite{zimmermann_semantics}, and there are similarly many ways to approach this topic for RDF streams. Therefore, we leave this topic for future discussion. The definitions are formulated in a semi-formal manner similar to the RDF 1.1 Concepts and Abstract Syntax W3C Recommendation~\cite{Cyganiak:14:RCA} or the RSP Data Model~\cite{rspmodel}, with the goal of making them readily applicable for RDF practitioners.

\subsection{RDF Stream Definitions}

We start by defining a universal, abstract RDF stream that will serve as a common denominator for further discussion:

\begin{Definition}
    An \textbf{RDF stream} is an ordered, potentially infinite sequence of RDF stream elements.
\end{Definition}
The manner in which the sequence is ordered is not specified on purpose, as this depends on the use case. Often, the order will be defined by time, but a stream does not have to be time-dependent (in fact, many methods do not consider time at all -- see Section~\ref{sec:literature}). This definition also does not specify what exactly are the RDF stream elements, as this is left to the further definitions that are based on it. 
Thus, we specialize the definition to specify this:

\begin{Definition}
    A \textbf{grouped RDF stream} is an RDF stream whose elements are either RDF graphs or RDF datasets.
\end{Definition}

\begin{Definition}
    A \textbf{flat RDF stream} is an RDF stream whose elements are statements (either RDF triples or RDF quads).
\end{Definition}
Grouped and flat RDF streams are still abstract, in the sense that their definitions are not concretized enough for practical implementation. However, they correspond to an important distinction between streams of groups of statements and streams of individual statements. With this, we can define the concrete types of RDF streams which can be used in practice:

\begin{Definition}
    An \textbf{RDF graph stream} is a grouped RDF stream whose elements are unnamed (default) RDF graphs.
\end{Definition}
In other words, an element of an RDF graph stream is a set of triples. This is a commonly used formulation that can be further specialized:

\begin{Definition}
    An \textbf{RDF subject graph stream} is an RDF graph stream in which every element contains an IRI node (called the subject node) that uniquely identifies the graph in the stream. Every other node in the graph can be reached by traversing triples, starting from the subject node.
\end{Definition}
RDF subject graph streams or similar formulations were proposed to handle timestamped RDF streams (TA-RDF~\cite{rodriguez2009semantic} and Stream Containers~\cite{schraudner2022stream}). ERI~\cite{fernandez2014efficient} uses a similar formulation for subject-molecules, but its subject IRIs are unique only within one block. The proposed definition is slightly broader than the timestamped models, while still being useful -- the advantage is that the element can be identified by an IRI, without using RDF datasets. The rest of the element can be discovered with what can be intuitively thought of as RDF graph connectivity.

We can then move on to streams of RDF datasets:

\begin{Definition}
    An \textbf{RDF dataset stream} is a grouped RDF stream whose elements are RDF datasets.
\end{Definition}
It should be noted here that although an RDF graph stream may seem like a special case of an RDF dataset stream, it is not. According to RDF 1.1, an RDF dataset is not a generalization of an RDF graph, but rather a collection of RDF graphs. Thus, the definitions of RDF dataset streams and RDF graph streams are not directly related in the taxonomy. It should be noted, however, that an RDF graph stream can be trivially transformed into an RDF dataset stream, by simply assuming that each graph in the former is the default graph in the dataset.

This definition can be further restricted to one named graph per element:

\begin{Definition}
    An \textbf{RDF named graph stream} is an RDF dataset stream in which every element has exactly one named RDF graph pair $\langle n, G \rangle$, where $G$ is an RDF graph, and $n$ is the graph name. Apart from graph $G$, the dataset may contain any number of triples in the default graph.
\end{Definition}
This is a simplified version of the RSP Data Model~\cite{rspmodel}, without the temporal aspect. Although we could find only one reference to this stream type in the literature (RSP engines can consume such streams if an appropriate conversion is implemented~\cite{rspmodel}), it is an intuitively important specialization of the RDF dataset stream. Here, every element corresponds to one named RDF graph with some optional information (metadata) about it in the default graph. As with RDF subject graph streams, this has the advantage of identifying a stream element by IRI, albeit with more flexibility as to the element's contents.

With this, we can move on to the timestamped variant:

\begin{Definition}
    \label{def:ts-named-graph}
    A \textbf{timestamped named graph} is an RDF dataset in which: \\ (1) there is exactly one named RDF graph pair $\langle n, G \rangle$, where $G$ is an RDF graph, and $n$ is the graph name; \\ (2) the default graph includes a timestamp triple $\langle n, p, t \rangle$, where $p$ is a timestamp predicate that relates $t$, called the timestamp, and the graph $G$.
\end{Definition}

\begin{Definition}
    \label{def:ts-named-graph-stream}
    A \textbf{timestamped RDF named graph stream} is an RDF named graph stream in which every element is a timestamped named graph. The elements that share the same timestamp predicate $p$ are ordered by the partial order associated with $p$.
\end{Definition}
Definitions~\ref{def:ts-named-graph} and~\ref{def:ts-named-graph-stream} were derived directly from the draft RSP Data Model~\cite{rspmodel}. Finally, we define concrete flat RDF stream types:

\begin{Definition}
    A \textbf{flat RDF triple stream} is a flat RDF stream whose elements are triples.
\end{Definition}

\begin{Definition}
    A \textbf{flat RDF quad stream} is a flat RDF stream whose elements are quads.
\end{Definition}
An ``RDF quad'' is a term that is often used by practitioners, but not defined in the RDF 1.1 Recommendation. It is however defined in the RDF 1.2 Working Draft (16 April 2024)~\cite{Hartig_rdf12}, and thus we use the RDF 1.2 draft definition of quads here.

Flat RDF streams can be represented simply as N-Triples or N-Quads files, however, contrary to RDF graphs and datasets, they have a defined order of statements. Similarly to RDF graph and dataset streams, a triple stream is not a subclass of a quad stream, because a triple is not a special case of a quad. However, a trivial transformation from one to the other can be performed, by stating explicitly that the triples belong to the default graph.

Grouped and flat RDF streams are notably interrelated, as any RDF graph stream can be flattened into an RDF triple stream, by simply listing and concatenating the statements in each of its elements -- the same is true for RDF dataset streams and RDF quad streams. The reverse is also possible, the elements of a flat stream can be grouped into a grouped stream. Such operations are in fact common in streaming applications~\cite{bonte2023streaming, fernandez2014efficient, keskisarkka2013semantic, sowinski2022efficient}.

Finally, for all of the above definitions, their generalized variants can be trivially derived (e.g., generalized RDF graph stream), by using generalized triples and generalized datasets as their basis. We skip these definitions for the sake of brevity.

\subsection{RDF Stream Taxonomy}
\label{sec:taxo}

\begin{figure}[htbp]
\centerline{\includegraphics[width=11cm]{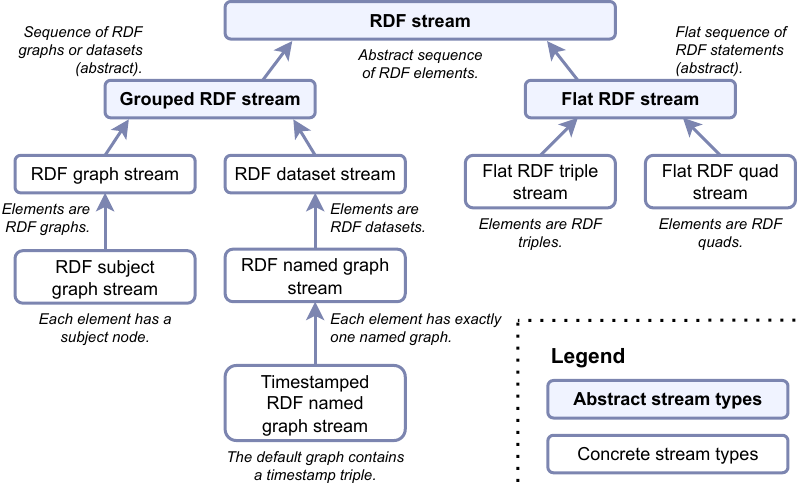}}
\caption{Overview of the RDF Stream Taxonomy (RDF-STaX).}
\label{fig:taxo}
\end{figure}

The proposed definitions form a taxonomical structure of task formulations (RDF Stream Taxonomy -- RDF-STaX), summarized in Figure~\ref{fig:taxo}. This overview allows us to observe that each concrete RDF stream type uses one of the four element types: graph, dataset, triple, and quad, possibly with additional restrictions on them. In fact, these four types of elements seem to be the only conceivable ones that can be derived from the RDF 1.1 specification. For each of them, representing the stream element is a known and solved problem (RDF serializations), and thus they can be easily used in practical applications. The abstract stream types are not directly usable in practice by themselves, but rather serve as the basis of the taxonomy and a shared starting point for other definitions.

\subsection{Taxonomy Correspondence to the State of the Art}
\label{sec:correspondence}

Finally, we link the proposed RDF stream types to the definitions used in state-of-the-art research and software -- Table~\ref{tab:mapping} presents such a summary. As noted in Section~\ref{sec:review_summary}, a single RDF stream can be viewed from multiple perspectives, which is reflected in the table, with some items appearing more than once when such a situation occurs. In some cases the correspondence is not ideal and there may be insignificant differences between the definitions proposed here and those in the literature. The short descriptions presented in the table are expanded upon in the nanopublications dataset, as described in Section~\ref{sec:nanopub}.

\begin{table}[htbp]
    \caption{RDF stream type usage in research and software, according to RDF-STaX.}
    \label{tab:mapping}
    \newcolumntype{s}{>{\hsize=.3\hsize}L}
    \begin{tabularx}{\linewidth}{sL}
        \toprule
        \textbf{Stream type} & \textbf{Uses}
        \\ \midrule
        
        RDF graph stream & 
        ActivityPub~\cite{Tallon:18:A},
        DBpedia-Live~\cite{morsey2012dbpedia},
        ERI (blocks)~\cite{fernandez2014efficient},
        GoT~\cite{le2016graph},
        IPSM (logical level)~\cite{ganzha2017streaming},
        Jelly (stream frames)~\cite{sowinski2022efficient},
        Keskis{\"a}rkk{\"a} \& Blomqvist~\cite{keskisarkka2013semantic},
        Live open scientific knowledge graphs~\cite{le2022towards},
        RDF EXI~\cite{kabisch2015exi}, 
        RDSZ~\cite{fernandez2014rdsz}, 
        RSP Data Model (can be consumed)$^{\dagger}$~\cite{rspmodel},
        RSP4J YASPER (can be consumed)~\cite{tommasini2021rsp4j},
        S-HDT~\cite{hasemann2012shdt},
        VoCaLS~\cite{tommasini2018vocals}
        \\ \midrule
        
        RDF subject\newline graph stream & 
        ERI (subject-molecule stream, only within a block)~\cite{fernandez2014efficient},
        LDES~\cite{van2021ldes},
        TA-RDF~\cite{rodriguez2009semantic},
        Stream Containers~\cite{schraudner2022stream}
        \\ \midrule
        
        RDF dataset stream & 
        IPSM (output)~\cite{ganzha2017streaming},
        Jelly (stream frames)~\cite{sowinski2022efficient}, 
        RMLStreamer~\cite{haesendonck2019parallel, oo2022rmlstreamer},
        RSP Data Model (can be consumed)$^{\dagger}$~\cite{rspmodel},
        Semantic Annotation enabler~\cite{semann}
        \\ \midrule
        
        RDF named\newline graph stream & 
        RSP Data Model (can be consumed)$^{\dagger}$~\cite{rspmodel}
        \\ \midrule
        
        Timestamped RDF\newline named graph stream & 
        Barbieri \& Della Valle$^{\ddagger}$~\cite{barbieri2010proposal},
        Linked Data Notifications for RDF streams~\cite{calbimonte2017linked},
        RSP Data Model~\cite{rspmodel},
        Tappolet \& Bernstein~\cite{tappolet2009applied},
        TripleWave~\cite{mauri2016triplewave},
        Wu et al.~\cite{wu2022workflow}
        \\ \midrule %\midrule
        
        Flat RDF\newline triple stream &
        Apache Jena RIOT~\cite{riot},
        ERI (high level)~\cite{fernandez2014efficient}, 
        Groppe et al.~\cite{groppe2007sparql},
        Jelly (high level)~\cite{sowinski2022efficient},
        Keskis{\"a}rkk{\"a} \& Blomqvist (mentioned)~\cite{keskisarkka2013semantic},
        RDF4J Rio~\cite{rio},
        RSP4J YASPER (can consume)~\cite{tommasini2021rsp4j},
        SPARQL-Generate~\cite{lefranccois2016flexible},
        VoCaLS~\cite{tommasini2018vocals}
        \\ \midrule
        
        Flat RDF\newline quad stream & 
        Apache Jena RIOT~\cite{riot}, 
        CARML~\cite{carml},
        Jelly (high level)~\cite{sowinski2022efficient}, 
        RDF4J Rio~\cite{rio} 
        \\ \bottomrule
        \multicolumn{2}{>{\hsize=\dimexpr1.3\hsize+1.3\tabcolsep+\arrayrulewidth\relax}X}{
            $^{\dagger}$ The RSP Data Model states that these stream types can be consumed by RSP engines by adding timestamps to each stream element. This is an optional feature of the RSP Data Model. \newline
            $^{\ddagger}$ The used representation differs from the definition in RDF-STaX by placing the metadata of the elements not in the default graph of the RDF dataset, but in a dedicated named graph.
        } \\
    \end{tabularx}
\end{table}

Different types of RDF streams appear to be applicable in different settings. In the context of network streaming applications, most commonly used are grouped RDF streams. Flat RDF streams, on the other hand, are used for streaming I/O and internally by various applications.

\section{RDF-STaX Ontology}
\label{sec:onto}

\begin{figure}[htbp]
\centerline{\includegraphics[width=11cm]{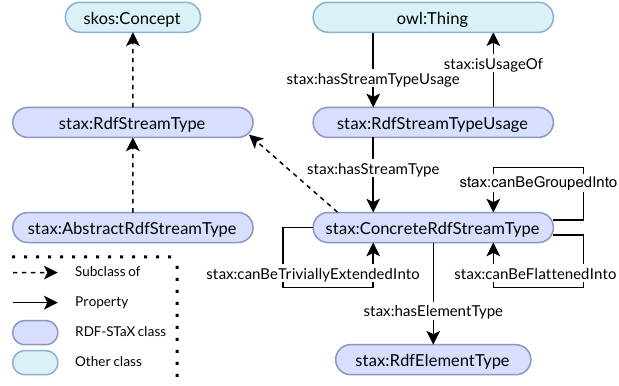}}
\caption{Classes and object properties in the RDF-STaX ontology.}
\label{fig:onto}
\end{figure}

To make the proposed taxonomy readily applicable, it was modeled as an OWL 2 DL ontology (Figure~\ref{fig:onto}), available at \url{https://w3id.org/stax/1.1.1/ontology} (accessed on 4 June 2024) under the Creative Commons Attribution 4.0 license. In the ontology, RDF stream types are instances of either the \texttt{stax:AbstractStream\-Type} or the \texttt{stax:ConcreteStream\-Type} class, with taxonomical relations being realized using SKOS in the OWL DL version~\cite{jupp2008skos, miles2005skos}. The lightest ontology profile that could be used for RDF-STaX is the unofficial OWL 2 DL, due to RDF-STaX being based on SKOS. This balances usefulness in the contexts where ontological reasoning is used (by being compatible with a number of reasoners~\cite{singh2020owl2bench}), with uses where ease of annotation takes precedence over semantics (e.g., in stream metadata or knowledge graphs with very little enforced semantics).

Each stream type instance in the RDF-STaX ontology is accompanied by a label, description, formal definition, and usage example. The instances are not shown in Figure~\ref{fig:onto}, but the structure they form is identical to the one presented in the taxonomy overview diagram (Figure~\ref{fig:taxo}), realized using the \texttt{skos:broader} property.

It should be stressed that RDF-STaX does not define a class for RDF streams, but rather classes for RDF stream \emph{types}. For example, \texttt{stax:namedGraphStream} is an instance of the \texttt{stax:ConcreteRdfStreamType} class. The ontology also defines a class for RDF stream type usages (\texttt{stax:RdfStreamTypeUsage}), which can be thought of as subjective views of how an RDF stream is used in a specific context. This design philosophy is closely reflected in the expected usage patterns, described in the next subsection.

The ontology includes rich semantic relations between stream types. Properties \texttt{stax:canBeFlattenedInto} and \texttt{stax:canBeGroupedInto} relate grouped and flat stream types that can be easily converted into one another. For example, an RDF graph stream can be flattened into a flat RDF triple stream by concatenating its elements, while a flat RDF quad stream can be grouped into a RDF dataset stream by punctuating it. The property \texttt{stax:canBeTriviallyExtendedInto} is used for graph- and triple-based streams, to indicate that they can be trivially turned into dataset- and quad-based streams, respectively. This transformation consists of stating explicitly that the triples are in the default graph. Object property chains are defined in the ontology to allow reasoning on these relations, exploiting the taxonomical relations from SKOS~\cite{miles2005skos}. These reasoning rules are presented below, in Manchester OWL Syntax~\cite{horridge2006manchester}:

\begin{minted}[breaklines]{turtle}
skos:broaderTransitive o stax:hasElementType
  SubPropertyOf: stax:hasElementType

skos:broaderTransitive o stax:canBeFlattenedInto
  SubPropertyOf: stax:canBeFlattenedInto

skos:broaderTransitive o stax:canBeGroupedInto
  SubPropertyOf: stax:canBeGroupedInto

skos:broaderTransitive o stax:canBeTriviallyExtendedInto
  SubPropertyOf: stax:canBeTriviallyExtendedInto
\end{minted}

The ontology in version 1.1.1 includes 5 classes, 8 object properties, 15 individuals, and a total of 192 axioms. The axiom count was obtained with the \texttt{measure} command in ROBOT 1.9.6~\cite{Jackson2019robot} and does not include the imported SKOS vocabulary. 
The \href{https://w3id.org/stax/1.1.1/ontology}{ontology's documentation} (accessed on 4 June 2024) contains a detailed description of every class, property, and individual. Additional information facilitating ontology reuse is included, such as an explanation of the available ontology formats, download links, and versioning information.

\subsection{Ontology Usage Patterns}

The proposed usage patterns for the RDF-STaX ontology reflect the philosophy of having a subjective view of how an RDF stream is used. These patterns are described in detail in the \href{https://w3id.org/stax/1.1.1/use-it}{ontology's documentation} (accessed on 4 June 2024), along with several illustrative examples and links to further resources.

The primary usage pattern in RDF-STaX is to create instances of the \texttt{stax:RdfStream\-TypeUsage} class to annotate research works, software, or datasets, with each instance corresponding to a different view or use case of the stream. This pattern was selected due to the inherent subjectivity of determining the types of RDF streams, as discussed in Section~\ref{sec:review_summary}. Below is an example of how this pattern may be used to annotate a DCAT~\cite{Browning:23:DCV} Dataset:

\begin{minted}[breaklines]{turtle}
@prefix dcat: <http://www.w3.org/ns/dcat#> .
@prefix rdfs: <http://www.w3.org/2000/01/rdf-schema#> .
@prefix stax: <https://w3id.org/stax/ontology#> .

_:dataset a dcat:Dataset ;
    # ... other properties of the dataset ...
    stax:hasStreamTypeUsage [
        a stax:RdfStreamTypeUsage ;
        stax:hasStreamType stax:datasetStream ;
        rdfs:comment "The data is a sequence of RDF datasets."@en
    ] , [
        a stax:RdfStreamTypeUsage ;
        stax:hasStreamType stax:flatQuadStream ;
        rdfs:comment "The data can be viewed as a flat sequence of RDF quads."@en
    ] .
\end{minted}
In some situations it may be more convenient to annotate existing resources from an external perspective. RDF-STaX also supports this, through the use of the \texttt{stax:isUsageOf} property. This is useful, for instance, when annotating published research papers or software with the RDF stream types that they use. This is illustrated in the example below, where the publication about IPSM~\cite{ganzha2017streaming} is annotated:

\begin{minted}[breaklines]{turtle}
@prefix rdfs: <http://www.w3.org/2000/01/rdf-schema#> .
@prefix stax: <https://w3id.org/stax/ontology#> .

_:usage a stax:RdfStreamTypeUsage ;
    rdfs:comment "The authors state that for their proposed system, 'the streaming element (i.e. a single message) (...) is a set of triples'. Therefore, internally it uses an RDF graph stream."@en ;
    stax:hasStreamType stax:graphStream ;
    stax:isUsageOf <https://doi.org/10.1109/ICSTCC.2017.8107003> .
\end{minted}

\subsection{Alignments to Other Ontologies}

The RDF-STaX ontology is aligned to other existing vocabularies with the use of semantic relations to foster the reuse of multiple ontologies linked together. Additionally, alignments can make it easier to understand the concepts introduced in RDF-STaX by comparing them to concepts already present in other vocabularies. However, as detailed in the comparison with the state of the art (Section~\ref{sec:onto_comp}), we could identify no other ontologies that would define the same concepts as RDF-STaX in a manner that would enable creating semantically strict, direct alignments (e.g., a subclassing relation). Instead, in cases where other ontologies define terms that are similar to those in RDF-STaX, a \texttt{skos:relatedMatch} alignment is provided. This essentially states that the meaning of the two concepts in different vocabularies is similar.

The following list enumerates the alignments that are provided in RDF-STaX release 1.1.1. Section~\ref{sec:onto_comp} contains further discussion on this topic, comparing the meaning of RDF-STaX's terms and those of other ontologies.

\begin{itemize}
    \item \textbf{SPARQL 1.1 Service Description vocabulary}~\cite{Williams:13:SSD} -- the classes for RDF graphs (\texttt{sd:Graph}) and datasets (\texttt{sd:Dataset}) were aligned with the corresponding terms in RDF-STaX (\texttt{stax:graph} and \texttt{stax:dataset}, respectively), which are instances of class \texttt{stax:RdfElementType}.
    \item \textbf{VoCaLS}~\cite{tommasini2018vocals} -- the \texttt{vocals:RDFStream} class was aligned with both the \texttt{stax:flat\-Triple\-Stream} and \texttt{stax:graphStream} instances in RDF-STaX.
    \item \textbf{LDES}~\cite{van2021ldes} -- the \texttt{ldes:EventStream} class was aligned with the \texttt{stax:subjectGraph\-Stream} instance in RDF-STaX.
    \item \textbf{VoID}~\cite{Zhao:11:DLD} -- the \texttt{void:Dataset} class was aligned with the \texttt{stax:flatTripleStream} instance in RDF-STaX.
\end{itemize}

\subsection{Availability and Sustainability}

The sources of the RDF-STaX ontology and its documentation are hosted on \href{https://github.com/RDF-STaX/rdf-stax.github.io}{GitHub} (accessed on 4 June 2024). The repository includes continuous integration and deployment (CI/CD) scripts that automatically validate the ontology's profile, run competency question tests (see Section~\ref{sec:app}), generate documentation, run an OWL 2 reasoner to add more assertions, package the ontology into a number of formats, and publish it on the \href{https://w3id.org/stax}{website} (accessed on 4 June 2024). Two versions of the ontology are published: OWL 2 DL, which can be used with reasoners (containing 249 triples); and OWL 2 Full (294 triples), which contains more metadata and alignments to other vocabularies. The latter is intended for use in contexts where ontological reasoning is not needed, e.g., for SPARQL querying. The additional metadata present there violates the OWL 2 DL profile, hence producing an OWL 2 Full ontology.

The ontology uses best practices for findability, accessibility, interoperability, and reusability (FAIR) developed by the community~\cite{garijo2020best}, compliance with which is evaluated in Section~\ref{sec:fair}. All releases of the ontology, including the OWL 2 DL and OWL 2 Full variants, are distributed through the RDF-STaX website, under a permanent URL hosted by the W3C Permanent Identifier Community Group. The ontology is also archived in Zenodo~\cite{zenodo} and registered in \href{https://archivo.dbpedia.org/info?o=https://w3id.org/stax/ontology}{DBpedia Archivo}~\cite{frey2020dbpedia} (accessed on 4 June 2024) and \href{https://lov.linkeddata.es/dataset/lov/vocabs/stax}{Linked Open Vocabularies}~\cite{vandenbussche2017linked} (accessed on 4 June 2024).

RDF-STaX is an open and freely-licensed project, welcoming contributions to the ontology, its documentation, and tooling. A \href{https://w3id.org/stax/1.1.1/use-it}{usage guide} (accessed on 4 June 2024) with several use case scenarios and examples was prepared to facilitate adoption. To ensure long-term sustainability, a comprehensive \href{https://w3id.org/stax/1.1.1/contributing}{contribution guide} (accessed on 4 June 2024) was provided, along with a \href{https://github.com/RDF-STaX/rdf-stax.github.io/issues}{public issue tracker} (accessed on 4 June 2024). The project is entirely hosted on the permanently free infrastructure of GitHub and w3id.org, which should ensure its long-term stability.

\section{Use Cases}
\label{sec:app}

In what follows, the three most representative use cases envisioned for RDF-STaX are described. The first use case is annotating published datasets and streams (for example, on the Web), to improve their accessibility and interoperability. The second use case focuses on achieving RDF streaming tool interoperability by embedding stream type metadata within the stream itself. The final envisioned application is analyzing the state of the art (SOTA) of RDF streaming in research and software, to easily compare different solutions and foster compatibility.

All of these use cases are described in detail below, along with accompanying competency questions (CQs) realized as SPARQL queries that should be run against the RDF-STaX ontology in the OWL 2 Full version. The CQ queries are available in \href{https://w3id.org/stax/1.1.1/uses/cq/}{RDF-STaX's documentation} (accessed on 4 June 2024) and are used to automatically and continuously test the ontology's validity with regard to the use cases, as described in Section~\ref{sec:uc_coverage}. Two of the presented use cases (published dataset annotation and analyzing the SOTA of RDF streaming) are already realized in practice, as detailed below. The three presented use cases should only be considered a starting point for RDF-STaX, as the community may over time find new ways in which to use it. The resource is expected to evolve over time (through a process called \emph{ontology evolution in the literature}~\cite{zablith2015ontology}) to remain up-to-date with the community's needs. To enable this, RDF-STaX adopts a thoroughly open paradigm, inviting outside contributions.

\subsection{Annotating Published Datasets and Streams}
\label{sec:uc-ann}

In this use case, RDF-STaX's stream types are used as metadata of datasets or streams published on the Web (for example, using the Linked Data mechanisms). This use case can be viewed from two sides -- the publisher's, and the end user's, which is reflected in the defined competency questions (see Table~\ref{tab:cq1}). The questions pertain to obtaining basic information about the defined stream types (CQ1.1 and CQ1.2). This can be used by the publisher to decide which stream type to use, and by the end user to understand what that stream type really is. Questions CQ1.3 and CQ1.5 play an analogous role, but also aim to provide additional context by relating the terms in RDF-STaX to external sources (other ontologies). Finally, CQ1.4 serves the end user by providing links to examples of how each stream type can be used in practice.

\begin{table}[htbp]
    \caption{Competency questions for use case 1 -- annotating published datasets and streams. Full SPARQL queries and expectations for each test can be found on the \href{https://w3id.org/stax/1.1.1/uses/cq/\#use-case-1}{ontology's website} (accessed on 4 June 2024).}
    \label{tab:cq1}
    \begin{tabularx}{\linewidth}{lL}
        \toprule
        \textbf{\#} & \textbf{Competency question}
        \\ \midrule
        CQ1.1 & What are the names and descriptions of all RDF stream types? \\
        CQ1.2 & What is the definition for each stream type? \\
        CQ1.3 & What is the type of element of each concrete stream type? Provide additional references to external sources for each element type, if available. \\
        CQ1.4 & How can each of the concrete stream types be used? Provide a link to one example for each. \\
        CQ1.5 & What are the corresponding terms from other ontologies to those defined in the RDF-STaX ontology? \\
        \bottomrule
    \end{tabularx}
\end{table}

This use case was implemented in RiverBench~\cite{sowinski2023riverbench}, an open RDF streaming benchmark suite, containing heterogeneous streaming datasets that are published using automated pipelines, following FAIR principles. RiverBench integrates with the RDF-STaX ontology, requiring all datasets to be annotated with their stream types. Each dataset has two stream types assigned, one for the grouped stream formulation and another for the flat. The consistency of the annotations is checked in CI/CD using SHACL rules~\cite{Knublauch:17:SCL} and the semantic relations defined in RDF-STaX (SKOS taxonomy properties and \texttt{stax:canBeFlattenedInto}). The datasets are then organized into benchmark profiles by their stream type. The published dataset documentation and RDF metadata based on DCAT~\cite{Browning:23:DCV} also includes the RDF stream type information.

At the time of submission, RiverBench's 2.0.0 release includes 12 datasets that use each stream type in RDF-STaX, except the RDF named graph stream (however, there are examples of the timestamped variant). The full metadata dump of RiverBench 2.0.0 consists of 9280 triples, including 201 \texttt{stax:hasStreamTypeUsage} annotations. Detailed technical information about this use case along with examples are available on the \href{https://w3id.org/stax/1.1.1/uses}{RDF-STaX website} (accessed on 4 June 2024).

\subsection{Embedding Metadata in Streams}

This use case focuses on embedding stream type metadata within the stream itself, with the goal of enabling streaming system interoperability. For example, a stream producer would state that the stream is of a given type, and then the consumer would check if it can consume this type of stream or convert it into a consumable stream. As this use case focuses on automated handling of streams, the competency questions (see Table~\ref{tab:cq2}) exploit the reasoning capabilities of RDF-STaX. Thus, CQ2.1, CQ2.2, CQ2.3, and CQ2.4 all address different possible transformations a stream processor could perform on an RDF stream.

\begin{table}[htbp]
    \caption{Competency questions for use case 2 -- embedding metadata in streams. Full SPARQL queries and expectations for each test can be found on the \href{https://w3id.org/stax/1.1.1/uses/cq/\#use-case-2}{ontology's website} (accessed on 4 June 2024).}
    \label{tab:cq2}
    \begin{tabularx}{\linewidth}{lL}
        \toprule
        \textbf{\#} & \textbf{Competency question}
        \\ \midrule
        CQ2.1 & Which concrete stream types can be viewed as generalizations of other stream types? \\
        CQ2.2 & Which concrete stream types can be trivially extended (by assuming the default graph) into other stream types? \\
        CQ2.3 & Which concrete stream types can be flattened into other stream types? \\
        CQ2.4 & Which concrete stream types can be grouped to obtain other stream types? \\
        \bottomrule
    \end{tabularx}
\end{table}

\subsection{Analyzing the State of the Art of RDF Streaming}
\label{sec:nanopub}

The final use case pertains to analyzing the SOTA of RDF streaming by annotating published research works and software with the RDF stream types used by them. This task is thus very closely related to the SOTA analysis performed in this paper, in Section~\ref{sec:correspondence}. As with any other scientific field, however, the landscape of RDF streaming changes rapidly and therefore using the RDF-STaX ontology in combination with a knowledge base of research works can help automate some literature review processes. Therefore, the defined competency questions (see Table~\ref{tab:cq3}) tackle more complex issues that may be asked when analyzing the correspondences between the annotated research works. CQ3.1 and CQ3.2 pertain to the taxonomical structure of RDF-STaX, helping the user understand how the different stream types relate to each other. Meanwhile, CQ3.3 attempts to find pairs of essentially incompatible RDF stream formulations, i.e., those that cannot be trivially converted into each other. This would serve to identify methods that cannot easily interoperate.

\begin{table}[htbp]
    \caption{Competency questions for use case 3 -- analyzing the state of the art of RDF streaming. Full SPARQL queries and expectations for each test can be found on the \href{https://w3id.org/stax/1.1.1/uses/cq/\#use-case-3}{ontology's website} (accessed on 4 June 2024).}
    \label{tab:cq3}
    \begin{tabularx}{\linewidth}{lL}
        \toprule
        \textbf{\#} & \textbf{Competency question}
        \\ \midrule
        CQ3.1 & What are the taxonomical parents of each stream type, listed in order? \\
        CQ3.2 & Is there any stream type that is a taxonomical parent or child of itself? \\
        CQ3.3 & Which conversions between concrete stream types cannot be done in any trivial way (grouping, flattening, extending)? Allow for multiple trivial transformations in series and take into account the taxonomical structure. \\
        \bottomrule
    \end{tabularx}
\end{table}

This use case was implemented within the scope of this work, with a nanopublication dataset of research works on RDF streaming. A nanopublication is a small unit of scientific knowledge published in RDF~\cite{kuhn2013broadening} -- it can be an opinion, a measurement, or any other assertion. Nanopublications are a natural way to express statements about scientific works or software due to their strong provenance mechanisms and citeability. This fits well the pattern of subjective statements about stream types, which is used in RDF-STaX. Therefore, to make it easier to create nanopublications with the RDF-STaX ontology, a template was prepared for Nanodash, an open nanopublication editor~\cite{kuhn2021semantic}. The template along with the \href{https://w3id.org/stax/1.1.1/nanopubs}{accompanying manual} (accessed on 4 June 2024) makes it easy to create and publish semantic assertions such as ``X uses stream type Y because Z'', requiring the user to only fill out a simple online form.

The results from the survey conducted in this paper (as summarized in Section~\ref{sec:correspondence}) were published as 35 nanopublications using the aforementioned mechanism, totaling 876 triples. The nanopublications include a more elaborated discussion of each stream type usage, providing more details than Table~\ref{tab:mapping}. The RDF-STaX nanopublications are packaged as a single knowledge graph in a CI/CD pipeline and published automatically on the \href{https://w3id.org/stax/1.1.1/nanopubs}{website} (accessed on 4 June 2024), where they can be downloaded and reused under the free CC BY 4.0 license. Anyone can contribute to the dataset by simply publishing nanopublications with the provided Nanodash template. Effectively, the dataset along with the publishing mechanism can serve as a collaborative, living state-of-the-art review, allowing one to easily compare and discuss works on RDF streaming, in a structured manner. A similar mechanism for living literature reviews was previously proposed by Wijkstra et al.~\cite{wijkstra2021living}, to whose work we refer the reader for more context.

\section{Ontology Evaluation}
\label{sec:evaluation}

This section presents the results of a multifaceted evaluation of the RDF-STaX ontology. Firstly, the evaluation includes checking use case coverage with the use of competency questions defined in Section~\ref{sec:app}. Secondly, the validity of the ontology is evaluated with regard to the OWL specification and its logical consistency. Thirdly, the findability, accessibility, interoperability, and reusability aspects of the ontology are evaluated, especially with regard to the current, community-developed best practices. Finally, RDF-STaX is compared with other similar resources, highlighting the advances offered by this work.

\subsection{Use Case Coverage}
\label{sec:uc_coverage}

This evaluation is based on the twelve competency question tests defined in Section~\ref{sec:app}. Each of these tests was written down as a SPARQL query paired with the expectation for the test, which specifies how many results should the query return when run against the RDF-STaX ontology. The tests are run in sequence by a CI/CD job implemented in the Python language, with the rdflib library~\cite{rdflib} -- source code is available on \href{https://github.com/RDF-STaX/ci-worker}{GitHub} (accessed on 4 June 2024) under the Apache 2.0 license. The job is triggered on every ontology release (including the development release) and pull request. If any test fails, the ontology will not be released, thus ensuring that it always covers the competency questions from the use cases. Further technical implementation details on the tests are available in the \href{https://w3id.org/stax/1.1.1/contributing/#competency-question-tests}{documentation} (accessed on 4 June 2024).

At the time of submission (RDF-STaX release 1.1.1), the CI/CD job reports that the ontology successfully passes all competency question tests. Additionally, the fact that the ontology was employed in two implemented applications (see Section~\ref{sec:app}) serves as additional validation of the ontology's usefulness and practicality~\cite{Sowiski2022quality}.

\subsection{Logical and OWL Profile Validity}

The purpose of this evaluation is to check if the ontology meets the requirements of specific OWL 2 profiles~\cite{Fokoue:12:OWO}, which strike different balances between the expressive power of the ontology and reasoning efficiency. Similarly to the use case coverage evaluation, the compliance checks with OWL 2 profiles were implemented as CI/CD jobs, which run on any ontology release and pull request. Both distributions of the ontology (OWL 2 DL and OWL 2 Full) are checked to see if they match their respective profile. The check is performed using ROBOT's~\cite{Jackson2019robot} \texttt{validate-profile} command. Additionally, the logical consistency of the ontology is validated in CI/CD with ROBOT's \texttt{reason} command, using the HermiT reasoner~\cite{Glimm2014hermit}.

At the time of submission (RDF-STaX release 1.1.1), the CI/CD job reports that both distributions of the ontology (OWL 2 DL and OWL 2 Full) are compliant with their respective profiles, with no inconsistencies. For additional validation, RDF-STaX 1.1.1 was checked with the publicly available Ontology Pitfall Scanner (OOPS!)~\cite{PovedaVillaln2014oops}, which reported no critical issues.

\subsection{Findability, Accessibility, Interoperability, and Reusability}
\label{sec:fair}

The question of what constitutes a FAIR ontology in practice is a complicated one~\cite{poveda2020coming}, with many possible interpretations for each of the FAIR Guiding Principles~\cite{Wilkinson2016}. To best reflect the needs of the Semantic Web community and ensure impartiality, in this evaluation we chose to rely on two well-known, publicly available, automatic FAIR evaluators: FOOPS!~\cite{garijo2021foops} and DBpedia Archivo~\cite{frey2020dbpedia}. The following results can be easily reproduced by inputting into either of them the RDF-STaX ontology IRI.

\begin{figure}[htb]
    \centering
    \begin{subfigure}[t]{\textwidth}
        \centering
        \includegraphics[width=\textwidth]{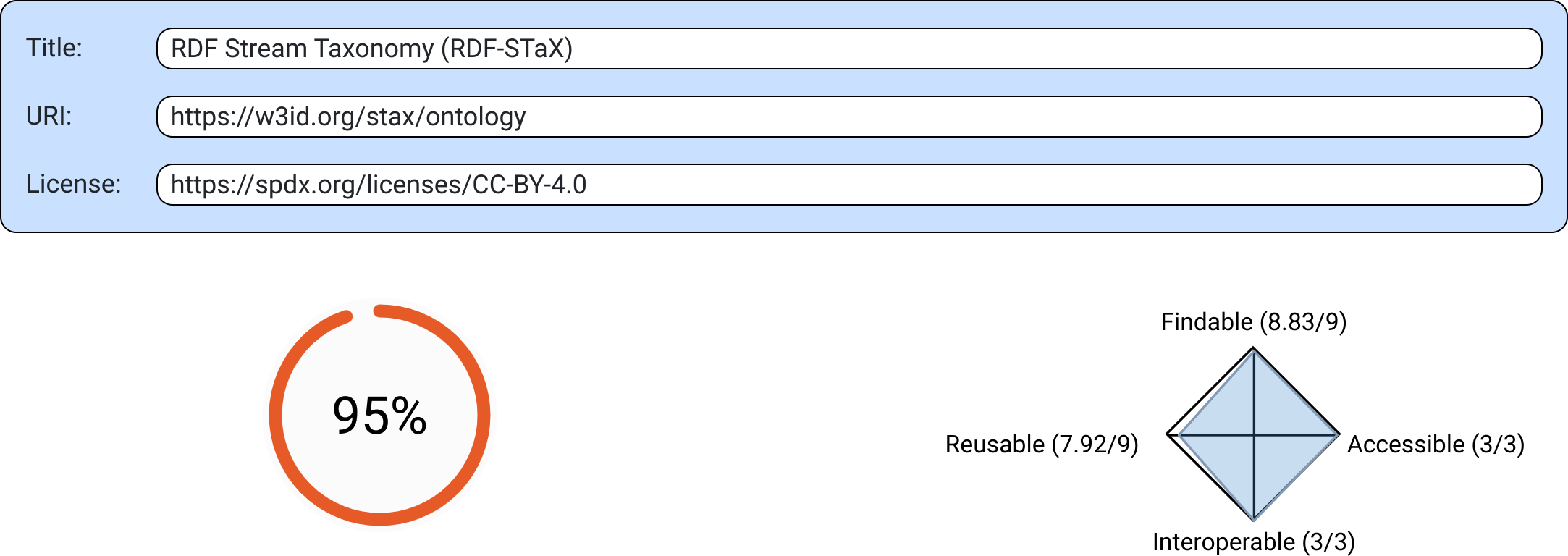}
        \subcaption{FAIR evaluation results obtained from FOOPS!~\cite{garijo2021foops}.}
        \label{fig:ext_eval_results_foops}
    \end{subfigure}
    \begin{subfigure}[t]{\textwidth}
        \centering
        \includegraphics[width=\textwidth]{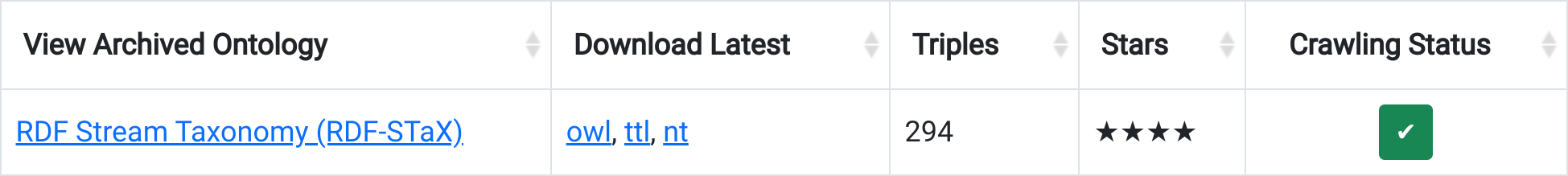}
        \subcaption{FAIR evaluation results obtained from DBpedia Archivo~\cite{frey2020dbpedia}.}
        \label{fig:ext_eval_results_archivo}
    \end{subfigure}
    \caption{FAIR evaluation results for RDF-STaX. Both screenshots were taken on 4 June 2024.}
    \label{fig:ext_eval_results}
\end{figure}

In FOOPS!, RDF-STaX obtained a score of 95\%, with 8.83/9 points in findability, 3/3 in accessibility, 3/3 in interoperability, and 7.92/9 in reusability (Figure~\ref{fig:ext_eval_results_foops}). All of the issues reported by the tool are minor. In DBpedia Archivo, RDF-STaX obtained 4/4 stars (Figure~\ref{fig:ext_eval_results_archivo}). It should be stressed here that Archivo's manual states that such a result means the ontology has reached ``minimum viability'' in the context of FAIR but is not necessarily ``good''. However, at the time of submission this is the highest possible rating offered by DBpedia Archivo, with only 20.03\% of all indexed ontologies having four out of four stars.

\subsection{Comparison to the State of the Art}
\label{sec:onto_comp}

To the best of our knowledge, there is no other ontology or vocabulary that would model RDF stream types. As previously noted, the RDF-STaX ontology purposefully does not define classes for RDF streams, but rather focuses on subjective RDF stream usages. This difference makes other ontologies non-overlapping -- instead, RDF-STaX can naturally complement them. As shown below, each existing ontology also covers only one or two selected formulations of the RDF streaming problem. Thus, RDF-STaX can provide a common semantic bridge to describe RDF stream types across different underlying vocabularies. To this end, the documentation includes \href{https://w3id.org/stax/1.1.1/use-it/}{examples} (accessed on 4 June 2024) of using RDF-STaX with these complementary ontologies (VoCaLS, VoID, LDES, and DCAT). The expected usage pattern is to define an RDF stream or a dataset with these ontologies as usual, and add the \texttt{stax:hasStreamTypeUsage} property to indicate the stream type according to RDF-STaX. The same pattern can be used with any future ontology for describing streams or datasets.

In what follows, the relevant ontologies are discussed in detail, focusing on their differences and complementarities with RDF-STaX. In the discussion, only publicly available ontologies are included, as otherwise it would not be possible to reliably assess the ontology's scope and semantics.

The VoCaLS vocabulary~\cite{tommasini2018vocals} is intended to help publish streaming data using the Linked Data principles and to describe streaming services. It defines the \texttt{vocals:RDFStream} class for potentially infinite sequences of RDF graphs and/or triples. Instances of this class can be thus annotated as RDF graph streams or flat RDF triple streams in RDF-STaX. As RDF-STaX does not define a class for RDF streams (but rather has instances of RDF stream types), the two ontologies are entirely complementary. Due to the different semantics of the terms in VoCaLS and RDF-STaX, a direct semantic connection (e.g., subclassing) between the two ontologies cannot be made. Instead, a \texttt{skos:relatedMatch} alignment is provided, along with a suggested usage pattern in the documentation of RDF-STaX.

The Linked Data Event Streams (LDES) vocabulary~\cite{van2021ldes} defines the \texttt{ldes:EventStream} class, which is a collection of stream elements -- sets of RDF triples. Each element is identified with its subject node. Therefore, instances of this class can be annotated in RDF-STaX as RDF subject graph streams. Such an alignment was added to the ontology.

The VoID vocabulary~\cite{Zhao:11:DLD} describes datasets with RDF data (not \emph{RDF datasets} as per RDF 1.1), which are in fact sets of triples with additional context information. Thus, instances of class \texttt{void:Dataset} can be annotated as RDF-STaX's flat RDF triple streams. This alignment was added to the ontology.

The more general DCAT~\cite{Browning:23:DCV} vocabulary describes datasets and their distributions. These datasets may be RDF streams, but DCAT does not assume any specific format of the datasets annotated with it. Therefore, RDF-STaX can be used as a complementary tool to annotate the type of the stream contained within such a dataset -- this is done for example in RiverBench, see Section~\ref{sec:uc-ann}.

The RDF-STaX ontology also defines four element types (RDF graphs, triples, quads, and datasets). In this regard, we could identify only the SPARQL 1.1 Service Description vocabulary~\cite{Williams:13:SSD}, which has the \texttt{sd:Graph} and \texttt{sd:Dataset} classes for RDF graphs and datasets. These classes were aligned with the corresponding terms in RDF-STaX, using the \texttt{skos:relatedMatch} property.

In summary, to the best of our knowledge, RDF-STaX is the first taxonomy and ontology that allows describing more than one type of RDF stream, making it a crucial step in enabling RDF streaming interoperability and understanding the field's state of the art.

\section{Discussion and Future Work}
\label{sec:discussion}

Although the ``RDF stream'' is a term that is now ubiquitous in research, it is used to describe many different concepts. There is no universally accepted language for describing RDF streams, which leads to a situation in which RDF streaming solutions become hard to assess, compare, and connect together. For example, forming a pipeline composed of a stream annotator, translator, streaming protocol, and a reasoner currently appears to be a tough challenge, as it is hard to determine what exactly is an RDF stream for each of these tools and whether these formulations are compatible with each other or not.

It appears that the variety of RDF stream formulations is here to stay. Each of the different definitions has its merits and use cases, and none of them should be disregarded. Instead, we call to embrace the variety of RDF streams by systematizing them in one shared taxonomy. The taxonomy with its semantic relations (taxonomical, flattening, grouping, extending -- see Section~\ref{sec:onto}) allows one to reason about stream types not only in terms of subclassing, but also in terms of possible conversions between them. This, we hope, will be a much needed step in the direction of making RDF stream applications more compatible with each other.

RDF-STaX is embodied in an ontology that makes the taxonomy readily usable in practice. The ontology does not attempt to replace any existing vocabulary for describing streams or datasets, but rather aims to complement them and serve as a shared semantic bridge between them. We hope that by being non-overlapping by design and focused on a single purpose, RDF-STaX will remain useful and relevant in the long term. The three presented use cases, two of which were already realized in practice, are common and well-known to the community, making the ontology all the more relevant.

Another result of this work is the nanopublications dataset, which is a collaborative, living state-of-the-art review of RDF streaming. Nanopublications, in our view, are a natural way to structure the intersubjective scientific discourse. It is possible for example, to comment on other nanopublications, confirm or disagree with them, all in a structured manner. Here we propose to use this powerful tool for a very narrow field of science (RDF streaming), but its possible future applications are certainly much broader~\cite{wijkstra2021living}. We encourage other researchers to contribute their perspectives to the dataset or to comment on the existing assertions.

As for future work, the theory presented in this contribution notably does not tackle the formal semantics of RDF streams. As with RDF datasets, it is unlikely that there can be single, unifying semantics for streams, but nonetheless, we see this area as worth exploring in the future. The presented survey of RDF streaming is also not a systematic review -- it only serves as a basis for the discussion in this work. Producing a systematic review of RDF streaming would be very valuable to the community, and we hope that the proposed taxonomy and the provided nanopublication tooling will make creating it easier.

\section{Conclusion}
\label{sec:conclusion}

In this work we propose the RDF Stream Taxonomy, which systematizes RDF stream definitions found in the literature. The taxonomy uncovered the variety of RDF streaming in research and practice, highlighting the open challenge of making the different applications compatible with each other. The proposed RDF-STaX ontology embodies the semantic relations between stream types in the taxonomy, and makes the definitions readily applicable. The ontology positions itself as complementary to existing vocabularies, and employs a flexible approach of making subjective statements about the types of RDF streams. The ontology and its documentation are fully open and follow community-developed best practices for FAIR.

The ontology was designed with three use cases in mind, with supporting competency questions for each. Two of these use cases were already implemented -- annotating research works in nanopublications, and describing RDF stream types of streaming datasets. The nanopublications dataset is an added contribution of this work, constituting a living state-of-the-art review that anyone can contribute to, using the provided documentation and tooling.

%%%%%%%%%%%%%%%%%%%%%%%%%%%%%%%%%%%%%%%%%%
\vspace{6pt} 

%%%%%%%%%%%%%%%%%%%%%%%%%%%%%%%%%%%%%%%%%%
%% optional

%%%%%%%%%%%%%%%%%%%%%%%%%%%%%%%%%%%%%%%%%%
\authorcontributions{Conceptualization, Piotr Sowiński, Paweł Szmeja and Maria Ganzha; Data curation, Piotr Sowiński; Formal analysis, Piotr Sowiński; Funding acquisition, Maria Ganzha and Marcin Paprzycki; Investigation, Piotr Sowiński; Methodology, Piotr Sowiński and Paweł Szmeja; Project administration, Maria Ganzha and Marcin Paprzycki; Resources, Maria Ganzha and Marcin Paprzycki; Software, Piotr Sowiński; Supervision, Maria Ganzha; Validation, Piotr Sowiński, Paweł Szmeja and Maria Ganzha; Visualization, Piotr Sowiński; Writing – original draft, Piotr Sowiński; Writing – review \& editing, Paweł Szmeja, Maria Ganzha and Marcin Paprzycki.}

\funding{This research received no external funding.}

\institutionalreview{Not applicable.}

\informedconsent{Not applicable.}

\dataavailability{The RDF-STaX ontology introduced in this study (along with the documentation and the nanopublications dataset) is openly available at \url{https://w3id.org/stax/1.1.1} (accessed on 4 June 2024) and in Zenodo at \url{https://zenodo.org/doi/10.5281/zenodo.10072907} (accessed on 4 June 2024), reference number 10072907.}

% Not applicable.
% \acknowledgments{In this section you can acknowledge any support given which is not covered by the author contribution or funding sections. This may include administrative and technical support, or donations in kind (e.g., materials used for experiments).}

\conflictsofinterest{The authors declare no conflicts of interest.} 

%%%%%%%%%%%%%%%%%%%%%%%%%%%%%%%%%%%%%%%%%%
%% Optional

\abbreviations{Abbreviations}{
The following abbreviations are used in this manuscript:\\

\noindent 
\begin{tabular}{@{}ll}
CI/CD & Continuous Integration and Continuous Delivery \\
CQ & Competency Question \\
DCAT & Data Catalog Vocabulary \\
FAIR & Findable, Accessible, Interoperable, Reusable \\
FOOPS! & Ontology Pitfall Scanner for FAIR \\
GoT & Graph of Things \\
I/O & Input/Output \\
IPSM & Inter Platform Semantic Mediator \\
IRI & Internationalized Resource Identifier \\
LDES & Linked Data Event Streams \\
OOPS! & Ontology Pitfall Scanner \\
OWL & Web Ontology Language \\
RDF & Resource Description Framework \\
RDF-STaX & RDF Stream Taxonomy \\
ROBOT & ROBOT is an OBO Tool \\
RML & RDF Mapping Language \\
RSP & RDF Stream Processing \\
SHACL & Shapes Constraint Language \\
SKOS & Simple Knowledge Organization System \\
SPARQL & SPARQL Protocol And RDF Query Language \\
TA-RDF & Time-Annotated RDF \\
URI & Uniform Resource Identifier \\
W3C & World Wide Web Consortium \\
\end{tabular}
}

%%%%%%%%%%%%%%%%%%%%%%%%%%%%%%%%%%%%%%%%%%
\begin{adjustwidth}{-\extralength}{0cm}

\reftitle{References}

\bibliography{bibliography}

%%%%%%%%%%%%%%%%%%%%%%%%%%%%%%%%%%%%%%%%%%
% \PublishersNote{}
\end{adjustwidth}
\end{document}